\documentclass{article}

\usepackage{amsthm}
\usepackage{graphicx}
\usepackage{float}
\usepackage{color}
\usepackage{amsmath}
\usepackage{amssymb}
\usepackage{multirow}
\usepackage{multicol}
\usepackage{authblk}

\title{Fixation in the stochastic Lotka-Volterra model with small fitness trade-offs}

\author[a]{Glenn Young}
\author[b]{Andrew Belmonte} 

 \affil[a]{Department of Mathematics, Kennesaw State University, Marietta, GA 30060}
\affil[b]{Department of Mathematics, Pennsylvania State University, University Park, PA 16802}

\begin{document}

\maketitle

\begin{abstract}
We study the probability of fixation 
in a stochastic two-species competition model. By identifying a naturally occurring fast timescale, we derive an approximation to the associated backward Kolmogorov equation that allows us to obtain an explicit closed form solution for the probability of fixation of either species. We use our result to study fitness tradeoff strategies and show that, despite some tradeoffs having nearly negligible effects on the corresponding deterministic dynamics, they can have large implications for the outcome of the stochastic system. 
\end{abstract}

\maketitle




Two species vying for the same ecological niche often have very small differences in competitive fitness \cite{Hubbell2005,Hubbell2006}. In such situations, small advantages can have large effects on survival outcome.
Of course, fitness advantages typically come at some corresponding cost to the organism. This principle is often referred to as a {\em tradeoff}: one fitness-increasing trait cannot be enhanced without another being diminished by the reallocation of resources \cite{Roff2007}. 
Population survival under competition can depend sensitively on such choices.

For instance, the bacterium {\em Salmonella Typhimurium} expresses a virulence factor that enhances its ability to displace the native commensal microbiota at the cost of reduced growth rate \cite{Sturm2011,Gog2012,Young2015}. 
The algae species {\em Chlorella vulgaris} similarly demonstrates a tradeoff between defense against grazing rotifers and its ability to compete for resources \cite{Yoshida2004,Jones2004}. Many more examples exist \cite{Nadell2011,Menge2008,Tarnita2015,Hereford2009,deFroment2014}, and in each example, one component of the species' competitive fitness is attenuated in favor of another. How can we understand the logic of these choices? A natural approach is by measuring the influence of these tradeoffs on the survival probability of the species.

Our present aim is to study such tradeoffs theoretically and tractably, building on recent mathematical tools. One possible method to this end is to incorporate tradeoffs into a deterministic ordinary different equations competition model. Unfortunately, such deterministic models struggle to capture the effects of small fitness changes, as the outcome depends continuously on model parameters and initial conditions, suggesting small changes in fitness-defining parameters should produce small changes in the dynamics. Consequently, this theory predicts that a small invading population with a small competitive advantage will never outcompete an established population. This result is at odds with empirical evidence \cite{Brown2006,Thiennimitr2011,Strogatz2018,Vaidyanathan2015,Wangersky1978,Young2015}.

Both classic and modern theoretical work in stochastic ecological modeling has focused on competition with a constant total population size, most notably using Wright-Fisher or Moran processes \cite{Hubbell1997,Hudson2002,Blythe2007,Danino2016,Schrempf2017,Shafiey2017,Danino2018,Danino2018b,Dean2017,Murray2020}. More recently, emphasis has been placed on models relaxing the assumption of constant population size,  describing ecological competition using Markov chains or stochastic differential equations \cite{Lin2012,Constable2015,Huang2015,Chotibut2017,Constable2016,Constable2018,Czuppon2018,Czuppon2018b}. Such stochastic models have the inherent advantage over their deterministic counterparts that the result of competition is determined only up to a probability distribution; that is, the outcome of competition is influenced, but not determined, by initial conditions. This allows for a more nuanced investigation into the effects of model parameters by allowing one to consider the relative change in the probability of success, rather than considering the outcome as a success-failure dichotomy. The probability that one population outcompetes the other, often called the probability of fixation, is therefore a central quantity measured to evaluate competitive strategies.  Because high-dimensional stochastic processes can be difficult, if not impossible, to analyze, methods of dimension reduction have emerged as important analytical tools. These methods often involve identifying slow and fast manifolds of the system, then collapsing the dynamics onto the slow manifold along the fast manifold \cite{Constable2013,Constable2016,Parsons2017}. 

Here we investigate the probability of fixation 
in a stochastic Lotka-Volterra competition model with a population that is allowed to fluctuate in size. We identify a natural slow timescale defined by a single model parameter, while exploiting the complementary fast time scale to derive an approximation to the backward Kolmogorov equation. This allows us to calculate an explicit expression for the probability of fixation as a function of the initial frequencies of the competitive populations. The simplicity of our approximation facilitates the efficient investigation of the effects of fitness differences on fixation probability and on the average time to extinction. 

We highlight the utility of our results by incorporating competitive tradeoffs into the model and studying the influence of these tradeoffs on fixation probabilities. In particular, we consider a trade off between growth rate and interspecific competition rate: a tradeoff observed within the invasion dynamics of {\em Salmonella Typhimurium} \cite{Sturm2011}. During invasion, {\em S. Typhimurium} maintain a subpopulation of an avirulent phenotype that grows more rapidly than the virulent counterparts, but does not express a virulence factor used to compete with the host's commensal microbiota. As the invasive bacteria begin to colonize the gut, the avirulent cells begin to express the virulence factor, allowing them to trigger the host's immune response to displace the commensal microbiota, but limiting their own growth rate. Our results offer an explanation of this behavior by showing that the tradeoff is beneficial when the {\em S. Typhimurium} population is small relative to the commensal microbiota population, but becomes harmful as the invading population becomes large. Through this illustrative example, we show that our results provide a quantification of the effects of small parameter changes in a competitive system that cannot be captured by deterministic dynamics.



\bigskip

We first consider the classic Lotka-Volterra competition model between two species $X(t)$ and $Y(t)$ as a continuous-time Markov process. We choose this model due to its ubiquity throughout theoretical competition literature \cite{Strogatz2018,Vaidyanathan2015,Wangersky1978,Young2015}, but note that the methods we develop in this work are not limited to this model, and can be applied to analyze other stochastic systems. The Lotka-Volterra model incorporates four modes of population dynamics: reproduction, natural death, death due to interspecific competition, and death due to intra-specific competition, sometimes referred to as crowding. Our goal is to quantify the effects of small fitness differences between the two competitive species on their respective probabilities of fixation; that is, the probability that one species goes extinct while the other persists. In particular, we will focus on differences in the parameters that define the two species' respective growth rates and their interspecific competition rates, while keeping all other fitness-defining parameters equal between the two populations. 


We consider these dynamics as a continuous-time Markov process defined by the transition probabilities
\begin{equation}\label{eq:markov}
\begin{aligned}
P^+_x(N_x,N_y)&=(f+\tilde{f}) N_x\\
P^-_x(N_x,N_y)&=d N_x + \frac{N_x(N_x-1)}{KM} + \frac{\alpha}{M} N_x N_y\\
P^+_y(N_x,N_y)&=f N_y\\
P^-_y(N_x,N_y)&=d N_y + \frac{N_y(N_y-1)}{KM} + \frac{\alpha+\tilde{\alpha}}{M} N_x N_y,
\end{aligned}
\end{equation}
where $P^\pm_s(N_x,N_y)$ denotes the transition probability rate of population $s\in\{X,Y\}$ to increase/decrease by one when $X=N_x$ and $Y=N_y$. The parameters $f$, $d$, $K$, and $\alpha$ define the baseline reproduction, natural death, intraspecific competition, and interspecific competition rates, respectively.  The parameter $M$ is assumed to be large and controls the stationary population size \cite{Huang2015,Czuppon2018}. The fitness difference between the two populations is defined by the two parameters $\tilde{f}$ and $\tilde{\alpha}$, which quantify the difference between the growth rate and the interspecific competition rate of the two populations, respectively. We will assume that these differences are generally small; in particular we will assume that $\tilde{f},\tilde{\alpha}\sim\mathcal{O}(1/M)$. Each parameter is assumed to be positive with the possible exceptions of $\tilde{f}$ and $\tilde{\alpha}$; $\tilde{f} > 0$ means that $X$ is more fit.

To facilitate mathematical analysis of the stochastic dynamics, we define the rescaled, approximately continuous variables $x=N_x/M$ and $y=N_y/M$. Reaction rates \eqref{eq:markov} expressed in terms of these new variables are given in Supplementary Information Section 1.

\begin{figure}[!b]
\centering 
 \includegraphics[width=1\linewidth]{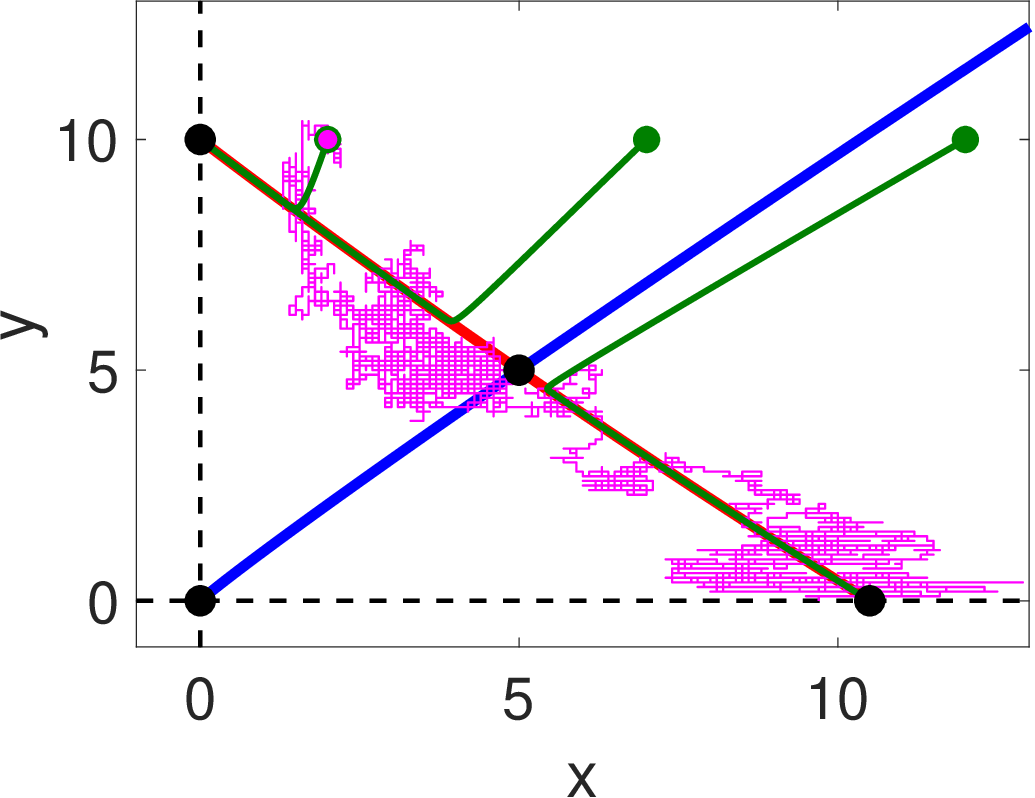}
\caption{Deterministic and stochastic solutions to the competitive Lotka-Volterra model in the population phase plane, with $f=1.1$, $\tilde f=0.05$, $d=0.1$, $\alpha=0.11$, $\tilde \alpha=-0.01$, $K=10$. Two solution trajectories (green curves) for the deterministic model Eqs.~\eqref{eq:detLV} start on the line defined by $y=(f-d)K$, quickly approach the unstable manifold (red curve), then move along the manifold toward one of the two stable equilibria. The dark blue curve is the stable manifold of the interior saddle point. 
The magenta curve shows a single realization of the Markov model \eqref{eq:markov}.
The black circles mark the four equilibria of the deterministic system and the green (magenta) circles mark the initial conditions of each trajectory. All stochastic simulations are implemented using a Gillespie algorithm \cite{Erban2007,Gillespie1977}.}
\label{fig:detLVPP}
\end{figure}

We now briefly consider the corresponding deterministic dynamics in order to determine viable parameter regimes. The Lotka-Volterra model for mutual competition is given by
\begin{equation}\label{eq:detLV}
\begin{aligned}
\dot{x}&=x\left(f+\tilde f-d-\frac{x}{K}-\alpha y\right)\\
\dot{y}&=y\left(f-d-\frac{y}{K}-(\alpha+\tilde \alpha) x\right),
\end{aligned}
\end{equation}
where the dot over the variables on the left hand side represents the derivative with respect to rescaled time $\tau=Mt$.  
%
System \eqref{eq:detLV} has four equilibria in general: the extinction state $(x,y)=(0,0)$, the $x$-only state $(x,y)=((f+\tilde f)K,0)$, the $y$-only state $(x,y)=(0,fK)$, and the coexistence state $(x^*,y^*)$, where $x^*$ and $y^*$ are functions of model parameters 
(see Fig \ref{fig:detLVPP}). We are presently interested in fixation probabilities, and therefore choose parameter values so that the deterministic model exhibits bistable behavior between two single-species equilibria; that is, one population will necessarily outcompete the other.  Straightforward linear stability analysis reveals that this is the case if  $\alpha(\alpha+\tilde \alpha)>1/K^2$; if this condition is satisfied, then the interior equilibrium $(x^*,y^*)$ is a saddle point while the two single-species steady states $((f+\tilde f)K,0)$ and $(0,fK)$ are both stable. We consequently will assume throughout this work that the inequality holds.

Fig \ref{fig:detLVPP} shows the phase plane of the deterministic system \eqref{eq:detLV}. Of course, the outcome of competition is entirely determined by the initial conditions in the deterministic model: solutions (green curves) that begin above the stable manifold (blue curve) of the saddle point result in population $y$ outcompeting $x$, while initial conditions below the same manifold result in $x$ outcompeting $y$. On the other hand, the magenta curve shows a single realization of the Markov model \eqref{eq:markov} that begins above the stable manifold of the interior saddle point of the deterministic system, yet results in $x$ outcompeting $y$: stochastic paths can cross the deterministic stable manifold.

\bigskip
  
In this bistable system, one of the two populations necessarily outcompetes the other, driving it to extinction. But in the stochastic model, the initial conditions merely influence probabilistically which population will win.
To quantify this, we study the probability that population $x$ outcompetes population $y$, referred to hereafter as the probability of fixation.  The central tool we use to determine the probability of fixation is the {\em backward Kolmogorov equation} (BKE; see Supplementary Information Equation 5). Given appropriate boundary conditions, steady state solutions of the BKE give the probability of fixation over varied initial conditions $(x_0,y_0)$ \cite{Ewens2012,Otto2011,Czuppon2018}. To ease calculation and interpretation, we follow \cite{Czuppon2018} and transform this BKE into the new coordinates 
$$
z(t)=x+y \qquad \text{and} \qquad p(t)=\frac{x}{x+y},
$$
which we call the demographic size and frequency, respectively. We rewrite the BKE in terms of $z$ and $p$ (see Supplementary Information, Eq 6).






\medskip


The steady state BKE of model \eqref{eq:markov} is a two-dimensional partial differential equation (Supplementary Information Section 2). Here, we construct a method of reducing the dimension from two to one, thereby converting the steady state BKE into an ordinary differential equation (ODE), allowing for much more efficient and tractable analysis. Recall that we are assuming that the fitness difference between the two populations is small, in particular that $\tilde f, \tilde \alpha \sim \mathcal{O}(1/M)$. 
We consequently write $\tilde f=\lambda/M$, $\tilde \alpha=-\mu/M$ (the minus sign in front of $\mu$ is included for future convenience), where $\lambda, \mu \sim \mathcal{O}(1)$. We further assume that $\alpha-1/K=\gamma/M>0$, where $\gamma\sim\mathcal{O}(1)$, requiring the difference between inter- and intraspecific competition rates be small and positive to remain in a bistable parameter regime. The parameter $1/M$ defines a natural length-scale in fitness space, and, with the above assumptions, the BKE is defined up to order $\mathcal{O}(1/M^2)$ (see Supplmental Material Section 2.2). Dropping the higher order terms, the BKE reduces to order $\mathcal{O}(1/M)$ as

\begin{equation}\label{eq:BKE}
\begin{aligned}
\frac{\partial u}{\partial t} &\approx \frac{p(1-p)}{M}\left[\lambda-\mu zp-\gamma z(1-2p)\right]u_p\\
&+\frac{p(1-p)}{2zM}\left[f+d+\alpha z\right]u_{pp}\\
&+z\left[f-d-\frac{z}{K}+\frac{1}{M}\left(\lambda p+\frac{1}{K}+zp(1-p)\left(\mu-2\gamma \right)\right)\right]u_z\\
&+\frac{z}{2M}\left[f+d+\frac{z}{K}\right]u_{zz},
\end{aligned}
\end{equation}
where $u=u(p,z)$ represents the probability of fixation with initial frequency $p$ and demographic size $z$.
 
We again consider the dynamics of the corresponding deterministic system. With the small-difference assumptions on model parameters above, the Eqs.~\eqref{eq:detLV} can be written in terms of $p$ and $z$ as follows:
\begin{equation}\label{eq:detLVpz}
\begin{aligned}
\dot{p}&=\frac{p(1-p)}{M}\left[\lambda-\mu p z-\gamma(1-2p)z\right]\\
\dot{z}&=z\left[f-d-\frac{z}{K}+\frac{1}{M}\left(\lambda p+\frac{1}{K}+zp(1-p)\left(\mu-2\gamma\right)\right)\right].
\end{aligned}
\end{equation}
The dynamics of system \eqref{eq:detLVpz} occur on two timescales: the fast dynamics of order $\mathcal{O}(1)$ and the slow dynamics of order $\mathcal{O}(1/M)$. Dropping the slow dynamics, the system reduces to the one-dimensional equation 
\begin{equation*}
\dot{z}=z\left(f-d-\frac{z}{K}\right).
\end{equation*} This indicates that the system approaches the manifold defined by $x+y=z=K(f-d)$ before competitive dynamics play a significant role (verified by simulations of System \eqref{eq:markov}; see Supplementary Information Fig S1), and once trajectories get close enough to this manifold, they remain close for all time. In particular, because the frequency variable $p$ remains approximately constant as the demographic variable approaches $z=K(f-d)$, the probability of fixation should not be dependent on the initial value of $z$, and should depend critically on the initial value of $p$.  We therefore make the ansatz that the probability of fixation does not depend on the initial value of $z$, and we consequently $u_z=u_{zz}=0$ (for similar methods, see \cite{Chotibut2017,Constable2018,Lin2012}). With this final approximation, the steady state BKE becomes
 
\begin{equation}\label{eq:ss}
\begin{aligned}
0 &= \frac{p(1-p)}{M}\left[\lambda-\mu pz-\gamma(1-2p)z\right]u'(p)\\
&+\frac{p(1-p)}{2zM}\left[f+d+\alpha z\right]u''(p).
\end{aligned}
\end{equation}
where $z=K(f-d)$. Solutions of this second order ODE subject to the boundary conditions $u(0)=0$ and $u(1)=1$ provide an approximation to the probability of fixation for species $X$.

Solving Eq \eqref{eq:ss} yields an explicit equation for the probability of fixation as a function of $p$:
%
%
\begin{equation}\label{eq:fixProb}
\begin{aligned}
u(p)=\frac{\displaystyle \int_0^p \exp\left(\frac{2z\left(-\lambda s+\mu z s^2/2+\gamma zs(1-s)\right)}{f+d+\alpha z}\right)ds}{\displaystyle \int_0^1 \exp\left(\frac{2z\left(-\lambda s+\mu z s^2/2+\gamma zs(1-s)\right)}{f+d+\alpha z}\right)ds}.
\end{aligned}
\end{equation}
This function can alternatively be written in terms of error functions (see Supplementary Information, Eq (10)).
%
%
A straightforward analysis of Eq.~\eqref{eq:fixProb} shows that, in the ``neutral'' case  
$\lambda=\mu=\gamma=0$, the probability of either population fixating is equal to its initial frequency: $u(p)=p$, in agreement with classic work \cite{Ewens2012,Czuppon2018}.

\begin{figure}[]
\centering
 \includegraphics[width=.49\linewidth]{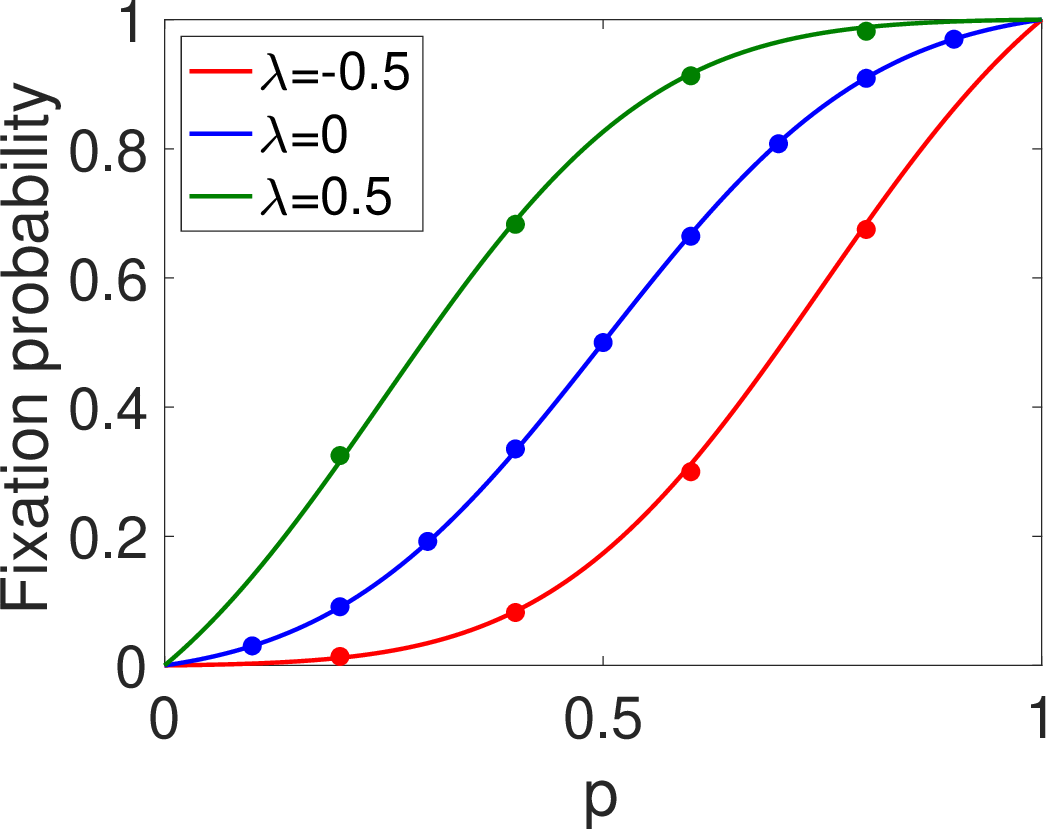} \includegraphics[width=.49\linewidth]{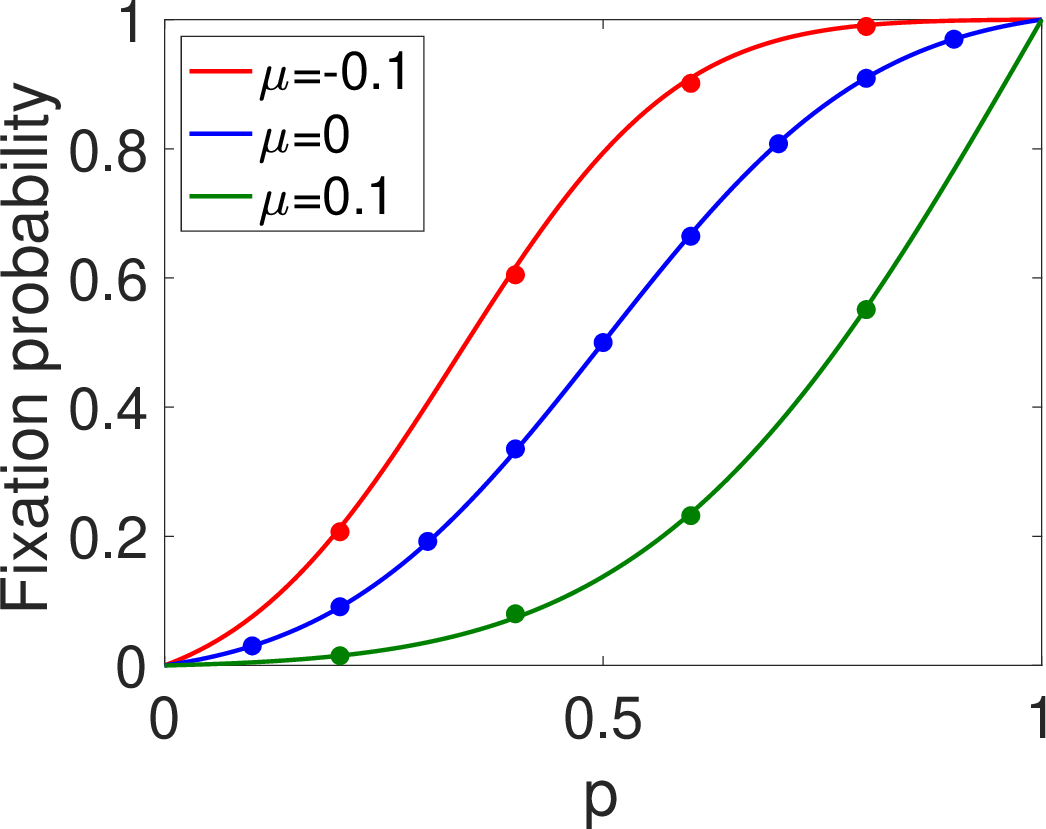}
\caption{Probability of fixation determined by the BKE, with $f=1.1$, $d=0.1$, $\alpha=0.11$, $K=10$, and $M=10$, showing: 
(L) variation with the enhanced growth rate $\lambda$, when $\mu=0$; 
(R) variation with enhanced competition rate $\mu$, when $\lambda=0$. The circles are numerical averages over 10,000 simulations of the Markov model \eqref{eq:markov}.
}
\label{fig:fixProbs1}
\end{figure}

More generally, we can use this formula to determine the influence of competition-defining parameters on the probability of fixation $u(p)$, 
 plotted for varied $\lambda$ and $\mu$ in Figure \ref{fig:fixProbs1}. Recall that $\lambda$ quantifies the increased growth rate in the $X$ population, and $\mu$ quantifies the decreased rate at which $X$ harms $Y$. Unsurprisingly, increasing $\lambda$ results in a uniform advantage for population $X$ for all initial proportions $p$, while increasing $\mu$ results in a uniform disadvantage for $X$. This can be shown in a straightforward manner in the $\gamma =0$ case by considering the second derivative of $u(p)$ (see Supplementary Information Section 3). Of course, these cases are clear: a uniform fitness advantage among one population will lead to a uniform increase in that population's probability of fixation. On the other hand, when a fitness advantage comes with an associated trade off, the influence on fixation probabilities is much less intuitively clear.

\bigskip

Our results facilitate the study of the effects of fitness tradeoffs on fixation probabilities. We focus on the tradeoff between growth rates and interspecific competition rates, which we couple by writing $\mu=\rho \lambda$ and varying $\lambda$. Increasing growth rate consequently comes at a cost of diminished competitive fitness, and vice versa. We couple these two parameters in order to narrow our focus to a single tradeoff, but emphasize that any two parameters can be similarly coupled. The parameter $\rho>0$ allows us to modulate the relative change between the two parameters, and we take some care in choosing its value. For $\rho$ very small or large, the resulting effects on the probability of fixation should closely match those in the uncoupled case (Figure \ref{fig:fixProbs1}). Specifically, for $\rho$ small, increasing $\lambda$ will result in an increase in the growth rate with little negative change in $\mu$, resulting in an overall competitive advantage for $X$. On the other hand, for $\rho$ large, increasing $\lambda$ will result in a substantial loss in the $X$ populations's competitive ability, causing a disadvantage for $X$.

For intermediate $\rho$, however, tradeoff effects become more complicated. In particular, we seek a value for $\rho$ so that increasing $\lambda$ from 0 does not uniformly increase or decrease the probability of fixation; that is, the tradeoff is advantageous for some range of $p$ and detrimental for another range. To do so, we differentiate $u(p)$ with respect to $\lambda$, and evaluate the result at $\lambda=0$. The sign of the resulting function over varied $\rho$ and $p$ determines for which values an increase in $\lambda$ will increase or decrease the probability of fixation. In particular, this determines a range of $\rho$ values so that the probability of fixation for small, positive $\lambda$ results in a higher probability for small $p$, but a lower probability for large $p$ (see Supplementary Information, Section 5).

Figure \ref{fig:fixProbs} shows the probability of fixation of population $X$ over initial proportion $p$ for varied $\lambda$, with $\mu=\rho\lambda$ and $\rho=0.2$. As the growth-rate advantage $\lambda$ increases, $X$ gains an advantage for small $p$ relative to $\lambda=0$, but loses this advantage for large $p$. 
This implies that advantageous competitive strategies depend on the initial frequencies. For example, a small, invasive population would benefit from this tradeoff, while a large, established population would suffer from it.

\begin{figure}[!b]
{
\centering 
\includegraphics[width=0.8\linewidth]{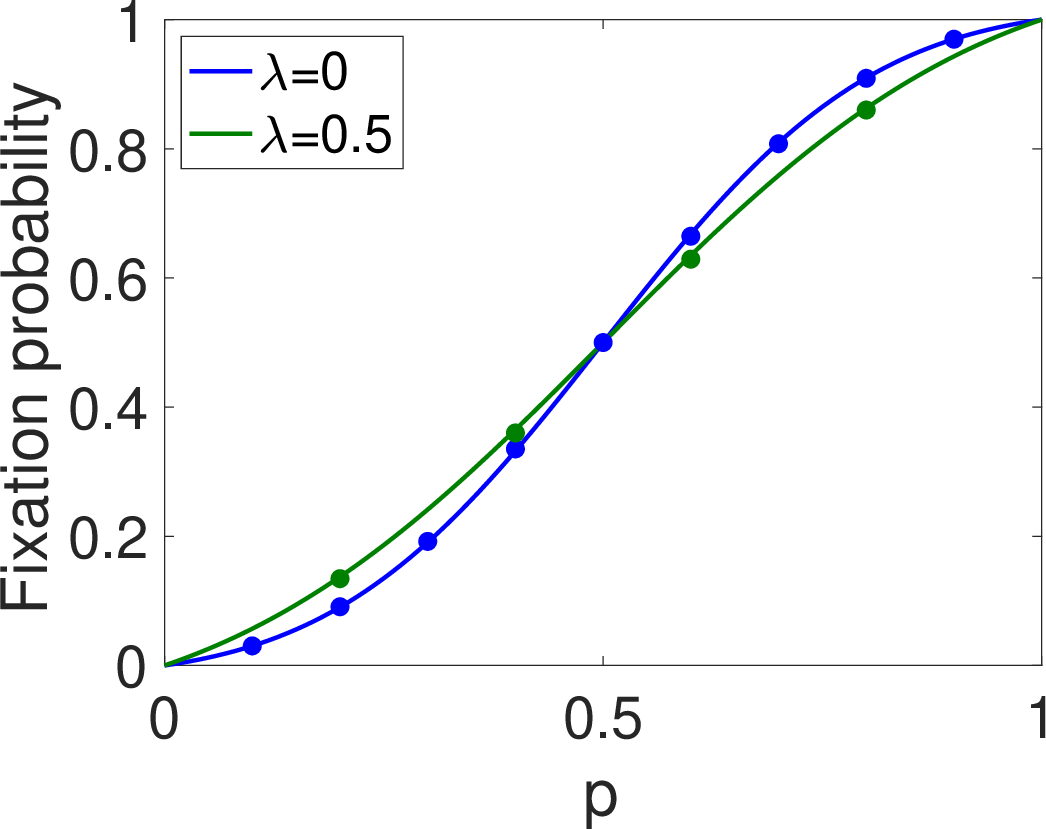}
}
\caption{Fixation probabilities 
over initial frequency $p$ with  
$\lambda=0$ and $\lambda=0.5$, using the tradeoff coupling $\mu=\rho \lambda$ with $\rho=0.2$, $f=1.1$, $d=0.1$, $\alpha=0.11$, $K=10$, and $M=10.$ The curves shown are Eq.~\eqref{eq:fixProb}, and the circles are numerical averages over 50,000 simulations of the Markov model \eqref{eq:markov}.}
\label{fig:fixProbs}
\end{figure}



Such density-dependent tradeoffs are known to play an important role in competitive processes. For example, the bacterial species {\em Salmonella Typhimurium} changes competitive strategies while invading the mammalian gut by modulating their expression of a virulence factor. These bacteria exploit the immune response of their hosts to displace native commensal microbiota from the intestinal wall, on which they colonize and grow \cite{Stecher2008,Santos2009,Sturm2011,Young2015}. Despite the competitive advantage conferred by virulence, {\em Salmonella Typhimurium} maintains 
 a subpopulation of avirulent cells that cannot trigger inflammation, but reproduce faster than their virulent counterparts \cite{Sturm2011}. These avirulent bacteria are able to activate the virulence factor and ``switch'' into the virulent population, and this activation rate increases over the late-logarithmic phase of growth \cite{Sturm2011}. 
In \cite{Young2015}, the authors find that {\em Salmonella Typhimurium} maximize their chance of outcompeting the commensal bacteria by invading with a large proportion of avirulent phenotypes. Their hypothesis is that the avirulent salmonella establish a ``foothold'' by reproducing quickly early, and then activate their virulence factor in order to trigger the host's inflammatory immune response, resulting in a higher interspecific competition rate against the commensal bacteria at a cost of diminished growth rate.  

While the stochastic Lotka-Volterra model considered here is only a crude approximation of the competitive dynamics between {\em Salmonella Typhimurium} and the commensal gut bacteria, our prediction qualitatively matches the above hypothesis: when there is a tradeoff between growth and competition rates, faster growth ($\lambda > 0$) is beneficial for small frequencies (lower $p$), while stronger competition is beneficial for larger frequencies (higher $p$).

We have not yet considered the mean time to fixation (or mean extinction time for the other population), which provides insight into rates of evolution and species lifetimes \cite{Danino2018,Frean2013}. This is often presented as an \textit{unconditional} average: that is, the average time until either species fixates, not a particular species. 
However, our derivation of an explicit formula for the probability of fixation also allows a straightforward, analytical method to find the \textit{conditional} mean time to fixation; for example, the average time it take population $X$ to fixate \textit{given that $X$ fixates} (see Supplemental Information, Section 4).


\medskip

Our derivation of an explicit formula approximating the probability of fixation in the competitive stochastic Lotka-Volterra model relies on two important observations. First, we identified a natural length-scale in fitness space, $1/M$, which we exploited to produce an approximation to the backward Kolmogorov equation. Second, we argued that the fast timescale of the dynamics governing the demographic variable, $z=x+y$, bring the system close to the manifold $z=K(f-d)$ before the frequency of population $X$ changes much from its initial state. This second observation allowed us to eliminate the fast $z$ variable, collapsing the backward Kolmogorov equation into a second order, one-dimensional partial differential equation, the steady state of which defines the probability of fixation. This probability of fixation offers a more careful interpretation of competitive strategies compared with analogous results from the corresponding deterministic systems of differential equations. Small fitness differences often have little effect on the dynamics of a deterministic system, which can greatly underestimate the importance of such differences. Our results allow us to quantitatively track the effect that small fitness differences induce on the probability of fixation.

The explicit formula derived from our approximation allows us to quantify the effects of fitness tradeoffs in competition. Such tradeoffs are common in nature and result from an inability to simultaneously optimize every fitness trait \cite{Roff2007}. Here we considered the tradeoff between growth rate and interspecific competition rate, using the tradeoff coupling $\mu=\rho \lambda$, however the same methods could be used to analyze tradeoffs between any number of fitness parameters, or for other tradeoff couplings $\mu=F(\lambda$). We showed the tradeoffs do not necessarily have a uniformly beneficial or detrimental effect on fixation probability. In particular, Fig \ref{fig:fixProbs} shows that fitness tradeoffs can be beneficial for some range of initial frequency $p$ but harmful for others. We argued that this is consistent with the invasion strategy used by {\em Salmonella Typhimurium}, in which the early presence of fast-growing avirulent phenotypes confers a competitive advantage over the host's native commensal bacteria, while a late increase in slow-growing virulent {\em Salmonella} triggers the host's immune response, differentially displacing the commensals, thereby increasing the interspecific competition rate of the {\em Salmonella} population.

The analysis and approximations we have presented are general enough to be applied to a range of problems in which a fast manifold exists and populations differ by small parameter perturbations. Similar recent work is considered in \cite{Constable2018}. As our approximation scheme was based on an ansatz, this approach should be made rigorous to determine error estimates. This can likely be done by adapting multi-timescaling methods from dynamical systems to our stochastic model \cite{Keener1995}.


{\bf Acknowledgments.}
This work was supported by National Science Foundation Grant CMMI-1463482.


\bibliographystyle{plain}
\bibliography{LVbib}

\end{document}